\documentstyle[12pt]{article}
\begin{document}
\setlength{\baselineskip}{24pt}

\begin{titlepage}
\centerline{\LARGE Density Matrix and Renromalization}
\centerline{\LARGE for Classical Lattice Models}
\centerline{Draft Ver. 1.0 (Sep. 9. 1996)}
\vskip 15pt
\centerline{ T. Nishino$^{1}$ and K. Okunishi$^{2}$ }
\centerline{\sl $1$ Department of Physics, Graduate School of Science, }
\centerline{\sl Kobe University, Rokko-dai, Kobe 657, Japan }
\centerline{\sl $2$ Department of Physics, Graduate  School of Science, }
\centerline{\sl Osaka University,  Toyonaka, Osaka 560, Japan }
\vskip 15pt

\begin{abstract}
\setlength{\baselineskip}{21pt}
The density matrix renormalization group is a variational approximation
method that maximizes the partition function --- or minimize the ground state
energy --- of quantum lattice systems. The variational relation is expressed
as  $Z = {\rm Tr} \, \rho \geq {\rm Tr} \, ({\tilde 1}\rho)$, where $\rho$ is the
density submatrix of the system, and ${\tilde 1}$ is a projection operator. In this
report we apply the variational relation to two-dimensional (2D) classical
lattice models, where the density submatrix $\rho$ is obtained as a product of
the corner transfer matrices. The obtained renormalization group method for 2D
classical lattice model, the corner transfer matrix renormalization group
method, is applied to the $q = 2 \sim 5$ Potts models. With the help of the finite
size scaling, critical exponents ($q = 2, 3$) and the latent heat ($q = 5$) are
precisely obtained.  \end{abstract}

\begin{itemize}
\item[\bf Address:] T.~Nishino, 
Department of Physics, Graduate School of Science,
\item[] Kobe university, Rokkodai, Kobe 657, JAPAN
\item[\bf Phone:] +81-78-803-0541, {\bf Fax:} +81-78-803-0722
\item[\bf e-mail 1:] nishino@phys560.phys.kobe-u.ac.jp
\end{itemize}
\end{titlepage}

\section{Introduction}

The basic procedure in the renormalization group (RG) is to keep relevant
information of a physical system, and neglect (or integrate out) irrelevant
one.~\cite{Kad,Wi,Rs} The density matrix renormalization group (DMRG)
introduced by White~\cite{Wh1} greatly enhances the applicability of
the numerical RG, because the method automatically keeps a fixed
numbers ($= m$) of the relevant basis; DMRG present the best
approximation within the limited numerical resource that we can use. The DMRG
has been applied to a number of one-dimensional (1D) quantum lattice systems,
such as the spin chain,~\cite{WF,SA} ladder,~\cite{Hida,WA} Bethe lattice
system,~\cite{Ot} strongly correlated electron
systems,~\cite{Yu,Noa,Shibata,Whtj} models in momentum space.~\cite{Xip} Not
only the numerical superiority, but also the formulation of DMRG have attracted
theoretical interests. \"Ostlund and Rommer~\cite{Os} have shown that DMRG is
a variational method, where the ground state is expressed as a product of
3-index tensors.~\cite{Zi,Ash} Mart\`\i n-Delgado and Sierra have investigated the
analytic formulation of DMRG, and have formulated the correlated block
RG.~\cite{Sierra,Si2} Recently White have refined the finite system algorithm
of DMRG, and extend the applicability of DMRG to 2D quantum
systems.~\cite{Wh2} Quite recently Xiang have reported DMRG study of 1D
quantum system at finite temperature,~\cite{Xift} using the quantum transfer
matrix formulation~\cite{QTM} and DMRG applied to the transfer
matrix.~\cite{Tmat}

Another RG approach has been development for 2D classical lattice models:
Baxter's method of the corner transfer matrix. (CTM)~\cite{Bax} The method is a
generalization of Kramers-Wannier approximation,~\cite{KrW,Kik} and
therefore Baxter's method is based on a variational principle for the partition
function. It should be noted that Baxter's variational relation is in principle the
same as the variational relation in DMRG.~\cite{CTMRG} The purpose of this
report is to explain how the concept of DMRG is applied to 2D classical lattice
models.  We start from a short review of the variational relation in DMRG in the
next section.

It is worth looking at a practical use of the RG method as the 2D photo image
compression.~\cite{JPEG,Kab} A photo image in our computer is normally
compressed before it is stored, in order to decrease the file size. The
compression algorithm is related to the block RG method,~\cite{Kad,Rs} since a
small region --- a pixel --- in a 2D picture has strong correlation with its
environment. (Computer scientists may insist that their findings about the
photo image compression are efficient for RG formulation in physics.) At
present, compression of movies --- TV pictures --- are in progress in the
world of computation;~\cite{MPEG1,MPEG2} the algorithm may be a good
reference for the RG study of 3D classical systems and 2D quantum systems. 

The development of RG, DMRG, Baxter's CTM method, and the photo
image compression is summarized in Fig.1. Originally these methods are
proposed independently, however, now it is apparent that their background is in
common. It is, to approximate a system (or the {\it objects}) within a limited
number of freedom.

\section{Variational Principle in DMRG}

We start from a short review of the variational principle in DMRG. We consider
the antiferromagnetic $S = 1/2$ Heisenberg spin chain as an example of 1D
quantum systems. The spin Hamiltonian is
\begin{equation}
H = J \sum_i {\bf S}_i \cdot {\bf S}_{i+1} ,
\end{equation}
where ${\bf S}_i$ represents the quantum spin at $i$-site, and the parameter
$J$ is positive. The Hamiltonian $H$ is real-symmetric, and so is the density
matrix $\rho = e^{-\beta H}$. (In the following discussion, $\rho$ does not
always have to be real-symmetric, but should be positive definite.)

We consider an open spin chain, (Fig.2) which consists of the left half [$L$] ($=$
the local system) and the right half [$R$] ($=$ the reserver). The terms `local
system' and `reserver' are rather formal, since [$R$] is not always longer than
[$L$]. (In Fig.2 both [$L$] and [$R$] has the same size.) The Hilbert space of the
whole system is spanned by the real-space basis $| \, l \, \rangle | \, r \rangle$,
where $| \, l \, \rangle$ and $| \, r \rangle$ corresponds to the spin configuration
for [$L$] and  [$R$], respectively. The matrix element of $\rho$ is given by
\begin{equation}
\rho_{l r, l' r'} = \langle \, l \, | \langle \, r | e^{-\beta H} | \, r' \rangle | \, l' \,
\rangle .
\end{equation}
In the context of DMRG, what is called the {\it density matrix} is actually the
density submatrix (DSM)
\begin{equation}
\rho^L = \sum_{l l'} | \, l \, \rangle \, \rho^L_{l l'} \, \langle \, l' \, |
\, \equiv \, \sum_{l l'} 
| \, l \, \rangle \bigg( \sum_r \rho_{l r, l' r} \bigg) \langle \, l' \, |
\end{equation}
that contains the information only about the local system [$L$]. The trace of
$\rho^L$ is equal to the partition function.

The relevant state selection --- renormalization --- in DMRG is performed
through the  diagonalization of the DSM
\begin{equation}
O^T \rho^L \, Q = {\rm diag}\{\lambda_1, \lambda_2, \ldots\} ,
\end{equation}
where $\lambda_1 \geq \lambda_2 \geq \ldots \geq 0$ are eigenvalues in
decreasing order, $Q = ({\bf q_1}, {\bf q_2}, \ldots)$ are the set of the
corresponding right eigenvectors, $O = ({\bf o_1}, {\bf o_2}, \ldots)$ are that of
the left eigenvectors. The matrices $O$ and $Q$ satisfies the dual orthogonal
relation $QO^T = 1$. When $\rho^L$ is real-symmetric, $Q$ is equal to $O$, and
both of them are orthogonal matrices. The first $m$ column vectors ${\bf
q_{\alpha}}$ ($1 \leq \alpha \leq m$) in $Q$ represent the RG transformation
from the original basis $\{ | \, l \, \rangle | \, r \rangle \}$ to the renormalized
basis $\{ | \, \alpha \, \rangle \}$. The irrelevant states are thrown 
away.~\cite{Wh1}

We can check the validity of the above basis state selection by observing
the inequality
\begin{equation}
Z = \sum_l \lambda_l \, \geq \, \sum_{l=1}^m \lambda_l .
\end{equation}
The quantity ${\tilde Z} = \sum_{l=1}^m \lambda_l$ is the approximate partition
function, which is smaller than $Z$ by $\sum_{l > m} \lambda_l$.
It has been known that if the
system has a finite excitation gap, the eigenvalue $\lambda_i$ decays
exponentially with respect to $i$, and therefore ${\tilde Z}$ is a good
approximation for $Z$ when $m$ is sufficiently large. The approximate partition
function is equal to the trace of the $m$-dimensional diagonal matrix 
\begin{equation}
{\tilde \rho}^L  
\equiv {\rm diag}\{\lambda_1, \lambda_2, \ldots, \lambda_m\}
= {\tilde O}^T \rho^L {\tilde Q} ,
\end{equation}
where ${\tilde Q}$ is the rectangular matrix $({\bf q_1}, {\bf q_2}, \ldots, {\bf
q_m})$, and ${\tilde O}$ is $({\bf o_1}, {\bf o_2}, \ldots, {\bf o_m})$. We can
regard the matrix operation of ${\tilde O}^T$ and ${\tilde Q}$ to $\rho^L$ as the
RG transformation (or the block spin transformation),
and ${\tilde \rho}^L$ as the renormalized DSM. 

Substituting Eq.(6) into Eq.(5), we obtain a variational relation in the matrix
form
\begin{equation}
Z = {\rm Tr} \, \rho^L \geq {\rm Tr} \, {\tilde \rho}^L 
= {\rm Tr} \, \big({\tilde Q}{\tilde O^T}\rho^L\big) ,
\end{equation}
where the matrix product ${\tilde 1} \equiv {\tilde Q}{\tilde O^T}$ has the
property of the projection operator $({\tilde Q}{\tilde O^T})^2 = {\tilde
Q}{\tilde O^T}$. The projection operator ${\tilde Q}{\tilde O^T}$ is {\it optimal} in
the sense that it gives maximum of ${\tilde Z} = {\rm Tr} \, ( {\tilde 1} \rho )$
under the constraint ${\tilde 1}^2 = {\tilde 1}$ and ${\rm Tr} \, {\tilde 1} = m$. In
other word, the DMRG minimize the free energy of the system within the
restricted degree of freedom. At the zero temperature ($\beta \rightarrow
\infty$), DMRG minimize the total energy.

The optimal projection operator ${\tilde 1}$ for the local system [$L$] is
dependent on the size of the reserver [$R$]. However, the dependence is
not conspicuous when [$R$] is sufficiently large. The infinite system DMRG
algorithm uses the insensitivity of ${\tilde 1}$ against the reserver size. The
finite system DMRG algorithm is more accurate than the infinite algorithm,
because the former correctly takes into account of the reserver-size
dependence.

\section{From Quantum system to Classical system}

In order to apply the variational relation in DMRG to 2D classical lattice system,
we define the DSM for 2D systems. We use the fact
that the density matrix $e^{-\beta H}$ of a $d$-dimensional quantum system can
be expressed as a partition function of a $d$+$1$-dimensional classical system
with special boundary conditions; the relation is known as the Trotter-Suzuki
formula.~\cite{Tr,Su}

The density matrix of the Heisenberg chain in Fig.2 is approximated as 
\begin{equation}
e^{-\beta H} = \bigg( e^{-{{\beta} \over {N}}H} \bigg)^M
\sim \bigg( e^{-{{\beta} \over {N}}{H_A}} e^{-{{\beta} \over {N}}{H_B}}\bigg)^M
\end{equation}
where $H_A \equiv \sum_i {\bf S}_{2i} \cdot {\bf S}_{2i+1}$ and 
$H_B \equiv \sum_i {\bf S}_{2i+1} \cdot {\bf S}_{2i+2}$ are the partition of
$H$. The matrix element $\rho_{l r, l' r'}$ in Eq.(2) is approximated by the
Boltzmann weight of the {\it chessboard model}, whose boundary spin
configurations are fixed to $l, r, l'$ and $r'$. Figure 3 shows an example when $M
= 4$. When we consider the partition function $Z$, the boundaries $l, r, l'$ and $r'$
plays the role of the tabs for sticking; $Z = {\rm Tr} \, e^{-\beta H}$ is
approximately equal to the partition function of the cylindrical 
system shown in Fig.4(a), which is constructed by attaching $l$ and $r$ in Fig.3 to
$l'$ and $r'$, respectively. In the same way, the approximate DSM $\rho^L_{l l'} =
\sum_r \rho_{l r, l' r}$ is obtained by  attaching $r$ to $r'$ as shown in Fig.4(b).

The expression of $\rho^L$ in Fig.4(b) is a typical example of the DSM for 2D
classical lattice models. The $\rho^L$ corresponds to the cylindrical system
with a cut $L$, where the spin configurations around the cut are fixed to $l$ and
$l'$. We generalize this example. Suppose we have a finite size lattice system
[$A$] shown in Fig.5. We then consider an arbitrary line or curve $L$ on  [$A$], and
cut [$A$] along $L$ to derive a new system [$A'$]. (The curve $L$ is a kind of
string on the $1$+$1$ space time.) The derived system [$A'$] has new boundaries
around the cut $L$, where the boundary spin configurations are represented by
the labels $l$ and $l'$. {\it The Boltzmann weight of} [$A'$], {\it which we write
$\rho^L_{l l'}$, is the DSM of} [$A$]. 

We have defined $\rho^L$ for 2D classical lattice models. Once we obtain
$\rho^L$ for a given system [$A$], we can perform DMRG along the variational
treatment discussed in the previous section, where the group of spins $l$ and
$l'$ on [$A$] are transformed into $m$-state effective spins. In the next section
we present an appropriate choice of [$A$] and $L$ for a typical 2D classical
model.

\section{Construction of the Density Matrix via CTM}

The Trotter-Suzuki decomposed Heisenberg spin chain in Eq.(8) is an anisotropic
2D lattice model, and the application of DMRG on this system is rather
complicated.\cite{Xift} For the tutorial purpose we consider 
a simpler 2D system, the symmetric 16-vertex model.~\cite{Bax,Miwa} The
model includes the Ising model~\cite{Is} as its special case. 
For simplicity, we assume that the model is ferromagnetic. 

The 16-vertex model is defined by the Boltzmann weight
\begin{equation}
W_{abcd} = W_{bcda} = W_{cdab} = W_{dabc}
\end{equation}
on each vertex ($=$ lattice point) of the simple square lattice, where the spin
variables $a, b, c,$ and $d$ take either $+$ (up) or $-$ (down). In the
following we consider a  square system [$A$], whose linear dimension is $2N$ or
$2N+1$. We impose fixed boundary condition on [$A$]; we introduce boundary
weights
\begin{equation}
P_{abc} = W_{abc+}
\end{equation}
and
\begin{equation}
C_{ab} = W_{ab++}
\end{equation}
in order to fix the boundary spins to $+$. (Fig.6) The partition function of [$A$]
is expressed by these weights. For example, the partition function of the
system with $2N+1 = 3$ is expressed as 
\begin{equation}
Z = \sum_{ab\ldots l} W_{kheb}^{~} P_{abc}^{~}  C_{cd}^{~} P_{def}^{~} 
C_{fg}^{~} P_{ghi}^{~} C_{ij}^{~} P_{jkl}^{~} C_{la}^{~},
\end{equation}
where the position of spin indices $a$-$d$ is shown in Fig.7.

In order to generalize Eq.(12) to arbitrary system size,
we introduce a half-row transfer matrix, (HRTM) which is a generalization of the
boundary weight $P$ in Eq.(10). The HRTM of length $N$ is defined by the
recursion relation  
\begin{equation}
P^N_{{\bf a}b{\bf c}} = \sum_d W_{a_N\,d\,c_N\,b}^{~} \, 
P^{N-1}_{{\bf a'}d\,{\bf c'}} ,
\end{equation}
where the label $N$ is the number of vertices in HRTM, 
$P^{\, 1}_{{\bf a}b{\bf c}}$ is equal to $P_{abc}$ in Eq.(10), and
${\bf a}$ represents a group of spins on a row
\begin{equation}
{\bf a} = (a_1, a_2,\ldots, a_{N-1}, a_N) ,
\end{equation}
which is related to ${\bf a'}$ as ${\bf a} = ({\bf a'}, a_N) $; the same for
${\bf c} = ({\bf c'}, c_N) $. Figure 8(a) shows an example when $N = 3$. We
occasionally drop the vector indices of $P^N_{{\bf a}b{\bf c}}$ and write it simply
as $P^N_b$; in that case we think of $P^N_b$ as a $2^N$-dimensional
matrix $(P^N_b)_{{\bf a}{\bf c}}$. The HRTM is also conventionally called {\it
vertex operator}.~\cite{Miwa2}

The density submatrix $\rho^L$ of the square system [$A$] is expressed as a
product of corner transfer matrices. (CTMs)~\cite{Bax} The CTM of size $N$ is
defined as 
\begin{equation}
C^N_{{\bf a}{\bf b}} = \sum_{{\bf c'}{\bf d'}} \bigg( 
\sum_{ef} W_{e\,f\,b_N\,a_N}^{~} \, 
P^{N-1}_{{\bf a'}e\,{\bf c'}} \, 
P^{N-1}_{{\bf b'}f\,{\bf d'}} \,  \bigg)
C^{N-1}_{{\bf c'}{\bf d'}} 
\end{equation}
where we have used the index rule ${\bf a} = ({\bf a'}, a_N)$, ${\bf b} =
({\bf b'}, b_N)$, etc. Figure 8(b) shows the example when $N = 2$. The smallest
CTM $C^1_{{\bf a}{\bf b}}$ is equal to the boundary weight $C_{ab}$ in Eq.(11).
The square system [$A$] is then constructed by attaching four CTMs. Figure 9
shows the system [$A'$] of the size $2N = 6$, whose Boltzmann weight
\begin{equation}
\rho^L = \big( C^N \big)^4 
\end{equation}
corresponds to the DSM of the square system [$A$] of the size $2N = 6$. Such a
construction of the DSM was first introduced by Baxter more than 30 years ago
in his variational method.~\cite{Bax} 

Now we can apply the RG procedure to 2D classical lattice models using $\rho^L$
defined in Eq.(16); following the RG procedure in Sec.2, we combine Baxter's
method and DMRG. For the brevity, we call our new RG method {\it corner
transfer matrix renormalization group} (CTMRG) in the following.~\cite{CTMRG}

\section{CTM Renormalization Group}

As was discussed in Sec.2, the heart of DMRG is the diagonalization of
$\rho^L$. For the symmetric 16-vertex model, where $\rho^L$
is expressed as  $\big( C^N \big)^4$,  both $\rho^L$ and $C^N$ have the
common eigenvectors. We therefore diagonalize CTM
\begin{equation}
O^T C^N Q = {\rm diag}\{\omega_1, \omega_2, \ldots\} 
\end{equation}
instead of $\rho^L$,
where we assume the decreasing order $|\omega_1| \geq |\omega_2|,
\ldots \geq 0$. The block-spin transformation is performed by the rectangular
matrices ${\tilde O} = ({\bf o}_1, {\bf o}_2, \ldots, {\bf o}_m )$ and ${\tilde Q} =
({\bf q}_1, {\bf q}_2, \ldots, {\bf q}_m )$, where ${\bf o}_{\alpha}$ and ${\bf
q}_{\alpha}$ are the left and the right eigenvectors of $C^N$, respectively.
Here after we use greek letters for indices that runs from $1$ to
$m$. (Actually, ${\tilde Q}$ is equal to ${\tilde O}$ since $C^N$ is symmetric.) The
renormalized CTM 
\begin{equation}
{\tilde C}^N \equiv {\tilde O}^T C^N {\tilde Q} 
= {\rm diag}\{\omega_1, \omega_2, \ldots, \omega_m \}
\end{equation}
is related to ${\tilde \rho}^L \equiv {\tilde
O}^T \rho^L \, {\tilde Q}$ via ${\tilde \rho}^L = ({\tilde C}^N)^4$.
The renormalized HRTM is obtained in the same way
\begin{equation}
{\tilde P}_a^N = {\tilde O}^T P_a^N {\tilde Q} ,
\end{equation}
where the tensor elements of ${\tilde P}_a^N$ are ${\tilde P}_{\xi a
\eta}^N$.

At this point we remember that $P^N$ and $C^N$ are defined through the
recursion relations Eq.(13) and Eq.(15), respectively. The relations are also valid
for ${\tilde P}^N$ and ${\tilde C}^N$. For ${\tilde C}^N$, its area is extended
by attaching two HRTMs (Fig.10) 
\begin{equation}
{\bar C}^{~N+1}_{(\alpha,a)(\beta,b)} 
= \sum_{ef \delta} W_{efba}^{~} \, {\tilde P}^N_{\alpha e \delta} 
{\tilde P}^N_{\beta f \delta} \, \omega_{\delta} ,
\end{equation}
where ${\bar C}^{~N+1}_{(\alpha,a)(\beta,b)}$ is a {\it partially renormalized}
CTM of linear size $N+1$, and the pairs of indices $(\alpha,a)$ and $(\beta,b)$
represent the row and the column matrix indices for ${\bar C}^{N+1}$. The
length of HRTM is simultaneously increased by putting a new vertex at the end
point
\begin{equation}
{\bar P}^{N+1}_{(\alpha,a) b (\gamma,c)} 
= \sum_d W_{adcb}^{~} {\tilde P}^N_{\alpha d \gamma}  \, .
\end{equation}

The extended CTM in Eq.(20) is not diagonal. As was done in Eqs.(17-18), we
diagonalize ${\bar C}^{N+1}$ to obtain the new renormalized CTM ${\tilde
C}^{N+1}$ with the matrix dimension $m$. It is now obvious that we can
indefinitely repeat the renormalization process in Eqs.(18-19) and  the system
size extension in Eqs.(20-21). In this way, we obtain ${\tilde P}^N$ and ${\tilde
C}^N$ for arbitrary large $N$ starting from $P^1$ and $C^1$.~\cite{CTMRG}

Approximate thermodynamic functions of [$A$] can be
obtained using ${\tilde C}^N$. The free energy $- {\sl k}_B T \,
{\rm ln} \, Z$ is estimated from the approximate partition function ${\tilde Z} =
{\rm Tr} \, \big({\tilde C}^N\big)^4 = \sum_i \omega_i^4$. It is also possible to
obtain local quantities, such as spin polarization and  multi-spin correlation
functions, using a combination of $P^N$ and $C^N$. For example, the
local energy is estimated as 
\begin{equation}
E(2N+1) = 
{ {\sum_{abcd} X_{abcd}^{~} {\rm Tr}
\big( {\tilde P}^N_a {\tilde C}^N {\tilde P}^N_b {\tilde C}^N 
{\tilde P}^N_c {\tilde C}^N {\tilde P}^N_d {\tilde C}^N \big) } \over
{\sum_{abcd} W_{abcd}^{~} {\rm Tr}
\big( {\tilde P}^N_a {\tilde C}^N {\tilde P}^N_b {\tilde C}^N 
{\tilde P}^N_c {\tilde C}^N {\tilde P}^N_d {\tilde C}^N \big) } },
\end{equation}
where the position of the indices $a$-$d$ are shown in Fig.11, and $X_{abcd}$ is
the local energy operator ${\rm ln} \, \big(W_{abcd}\big)$.  Since the CTMRG
extends the area of the system [$A$] from the center, local quantities at the
center can be calculated most precisely.

The largest matrix element of ${\tilde C}^N$ rapidly grows with respect to $N$.
We should therefore normalize CTM
\begin{equation}
{{{\tilde C}^N} \over {\omega_1}} = {\rm diag}\bigg\{ 1, 
{{\omega_2} \over {\omega_1}}, \ldots,  
{{\omega_m} \over {\omega_1}}\bigg\} \rightarrow {\tilde C}^N
\end{equation}
in realistic numerical calculations. We should also normalize ${\tilde P}_a^N$ in
the same manner. Apart from the critical point, the normalized ${\tilde C}^N$
converges to its thermodynamic limit ${\tilde C}^{\infty}$ exponentially with
respect to $N$. At criticality the convergence is relatively slow; it is observed
that the decay rate at criticality is controlled by the critical
exponents.~\cite{NOK}

We have imposed fixed boundary conditions on [$A$]. Since
the boundary condition is totally determined by the boundary weight $P_a^1$ in
Eq.(10) and $C^1$ in Eq.(11), we can choose other boundary conditions by
modifying $P_a^1$ and $C^1$. For example, the free boundary condition is 
imposed by the boundary weights
\begin{equation}
P_{abc} = W_{abc+} \, + \, W_{abc-}
\end{equation}
and
\begin{equation}
C_{ab} = W_{ab++} \, + \, W_{ab+-} \, + \, W_{ab-+} \, + \, W_{ab--} .
\end{equation}

The CTMRG method presented above can be applied to a wide class of 2D
classical lattice models, such as the $q$-state Potts model, IRF model, etc. We
have to be careful to anisotropic lattice models whose  $\rho^L$ consist of four
{\it different} CTM's. (In Baxter's textbook the DSM are expressed as $\rho =
ABCD$.~\cite{Bax}) In such a case, we should not diagonalize each CTM
independently, but we should diagonalize $\rho^L$ to perform RG
transformation. We should also be careful to antiferromagnetic models, because
we have to prepare several sets of ${\tilde P}^N$ and ${\tilde C}^N$ according to
the alternating spin order.

\section{Numerical Result}

The symmetric 16-vertex model includes the square lattice Ising model as its
parameter limit. We first look at the critical phenomena of the Ising model
using CTMRG. Figure 12 shows the calculated local energy
\begin{equation}
E(2N) = {\rm Tr} \, \big( \sigma \sigma' {\tilde \rho}^L \big) / {\tilde Z}
\end{equation}
where $\sigma \sigma'$ is a pair of neighboring spin at the center of the square
cluster [$A$] of size $2N$. The data shown by the black dots are
$E(\infty)$, that are obtained when $m = 98$. The data deviate from the exact
solution at most $10^{-7}$. At the critical temperature $T_c$, we estimate
$E_c(\infty)$ by observing its convergence with respect to $N$. The inset of
Fig.11 shows the $1/N$ dependence of $E_c(N)$. A simple $1/N$
fitting gives $E_c(\infty) = 0.707148$, which is close to the exact one $1 /
\sqrt{2} = 0.707107\ldots$.

The $1/N$ dependence of $E_c(N)$ is actually related to the scale invariance at
criticality. The finite size scaling (FSS) theory~\cite{Fi,Ba} predicts that
$E_c(N)$ obeys the scaling form
\begin{equation}
E_c(N) - E_c(\infty) \sim N^{1/\nu -d},
\end{equation}
where $\nu$ is the correlation-length exponent and $d =2$ is the spatial
dimensionality; in our case $E_c(N)$ is proportional to $N^{-1}$ because $\nu$ of
the Ising model is equal to unity. Similarly, the local order parameter 
\begin{equation}
M(N) = {\rm Tr} \, \big( \sigma {\tilde \rho}^L \big) / {\tilde Z}
\end{equation}
at the center of [$A$] obeys
\begin{equation}
M_c(N) \sim N^{-(d-2+\eta)/2} 
\end{equation}
with the anomalous dimension of the spin $\eta$. Figure 13 shows the $N$
dependence of calculated magnetization $M_c(N)$ at $T_c$ when $m = 148$.
As it is expected from $\eta = 1/4$, the calculated $M_c(N)$ is proportional to
$N^{-1/8}$.

The estimation of the critical exponents via FSS is valid for a wide class of
2D classical model. As examples, we apply CTMRG and FSS to the $q$=$2, 3$
Potts models,~\cite{Pt,Wu} which can be treated as a symmetric $q^4$-vertex
model.~\cite{NO2,Wu} Table I shows the estimated exponents $\nu$ and $\eta$
from the calculated data when $m = 200$ and $100 \leq N \leq 1000$. For
comparison, the theoretically determined exponents~\cite{Wu} are shown in the
parenthesis. The error in the calculated exponents are less than 0.2\%.

\begin{table}
\caption{Critical exponents $\nu$ and $\eta$ of the $q$=$2, 3$
Potts model estimated from the numerical data when $m = 200$. Theoretical
values are shown inside the parenthesis. } 

\begin{tabular}{@{\hspace{\tabcolsep}\extracolsep{\fill}}ccc} \hline
q &  $\nu$ (Exact) & $\eta$ (Exact) \\ \hline
2 &  1.0006 (1.0000)  & 0.2501 (0.2500) \\
3 &  0.8321 (0.8333)  & 0.2654 (0.2667) \\ \hline
\end{tabular}
\end{table}

Numerical analysis of the case $q = 4$ is in progress. In this case,
one has to take care of the logarithmic corrections in the FSS
analysis.~\cite{q4}

We finally show the $N$ dependence of the local energy $E(N)$ of the $q$=$5$
Potts model at the transition temperature; we measure $E(N)$ from the average
$E^* = (E_{+} + E_{-})/2$, where $E_{+}$ and $E_{-}$ are the energy of
the disordered and the ordered phase, respectively, at the transition
temperature. (Fig.14) Since the transition is first order, $E(N) - E^*$ does not
obey the scaling form in Eq.(27). In this case, it is possible to estimate the
latent heat through the scaling analyses with respect to both $N$ and $m$. The
calculated latent heat is $0.0256$, where the exact value is $0.0265$.~\cite{Wu}
The CTMRG is expected to be efficient for the analysis of the weak first order
transition.~\cite{q5}

\section{Conclusion and discussion}

We have reviewed the variational principle in DMRG, and explained its
application to 2D classical lattice systems. We create the DSM according
to Baxter's construction, and perform the RG transformation
using DMRG algorithm. As trial calculations, we perform FSS analyses on
the $q$-state Potts models at the transition temperature. It is concluded that
the CTMRG is efficient for the determination of the critical exponents or the
latent heat.

It is possible to feed-back the CTMRG algorithm to the zero-temperature
1D quantum system, and to obtain a rapid infinite system DMRG
algorithm; the product wave function (PWF) RG.~\cite{PWF} The PWFRG is
closely related to the improved finite system DMRG algorithm (DMRG++ in
Fig.1) recently proposed by White.~\cite{Wh2} 

Finally, we discuss the applicability of DMRG or CTMRG to 3D classical
lattice models. As we have employed a square system for  2D models, we should
consider a cubic system for 3D models. We divide it into 8 subcubics, say corner
transfer tensor, (CTT) and construct the DSM as their tensor products. We then
diagonalize the DSM, and renormalize CTT.  Apart from the renormalized CTM in
two dimension, renormalized CTT is not diagonal.~\cite{BaxZ} There is,
however, a computational problem that prevents us from practical use of the RG
method for 3D system; at present the numerical calculation is too heavy even
for a small $m$. 

There are plenty of subject that we can clarify in the field of the numerical RG.

\section*{Acknowledgments}

The authors would like to express their sincere thanks to Y.~Akutsu and
M.~Kikuchi for valuable discussions. T.~N. thank to G.~Sierra, M.~A.~Mart\`\i
n-Delgado  and S.~R.~White for helpful discussions about the RG method. The
numerical result on $q = 5$ Potts model is obtained through the
collaboration with A.~Yamagata, H.~Otsuka and Y.~Kato.~\cite{q5} The present
work is partially supported by Kasuya foundation. Most of the numerical
calculations were done by NEC SX-3/14R in computer center of Osaka
University.

\newpage

\section*{Figure Captions}

\begin{itemize}

\item[\bf Fig.~1.] Historical overview of DMRG and related fields.

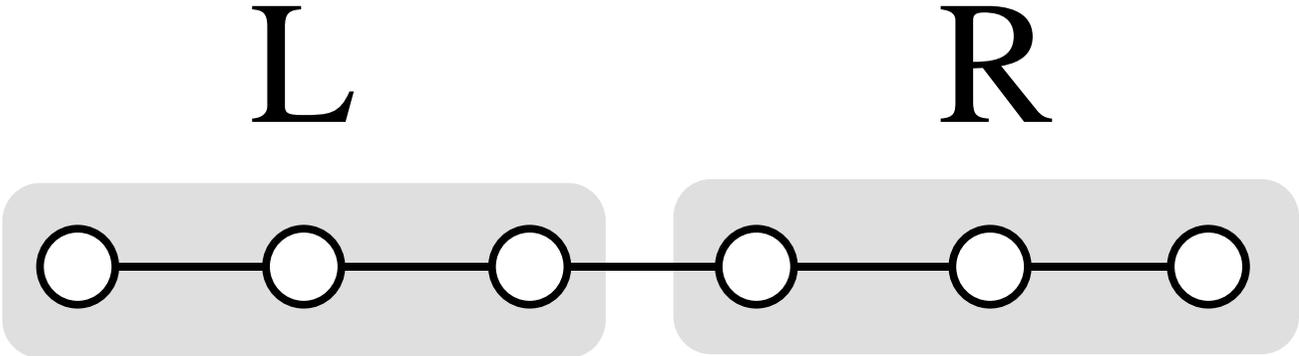
\item[\bf Fig.~2.] The $S = 1/2$ Heisenberg spin chain with open boundary
conditions. We divide it into the local system [$L$] and the reserver [$R$].

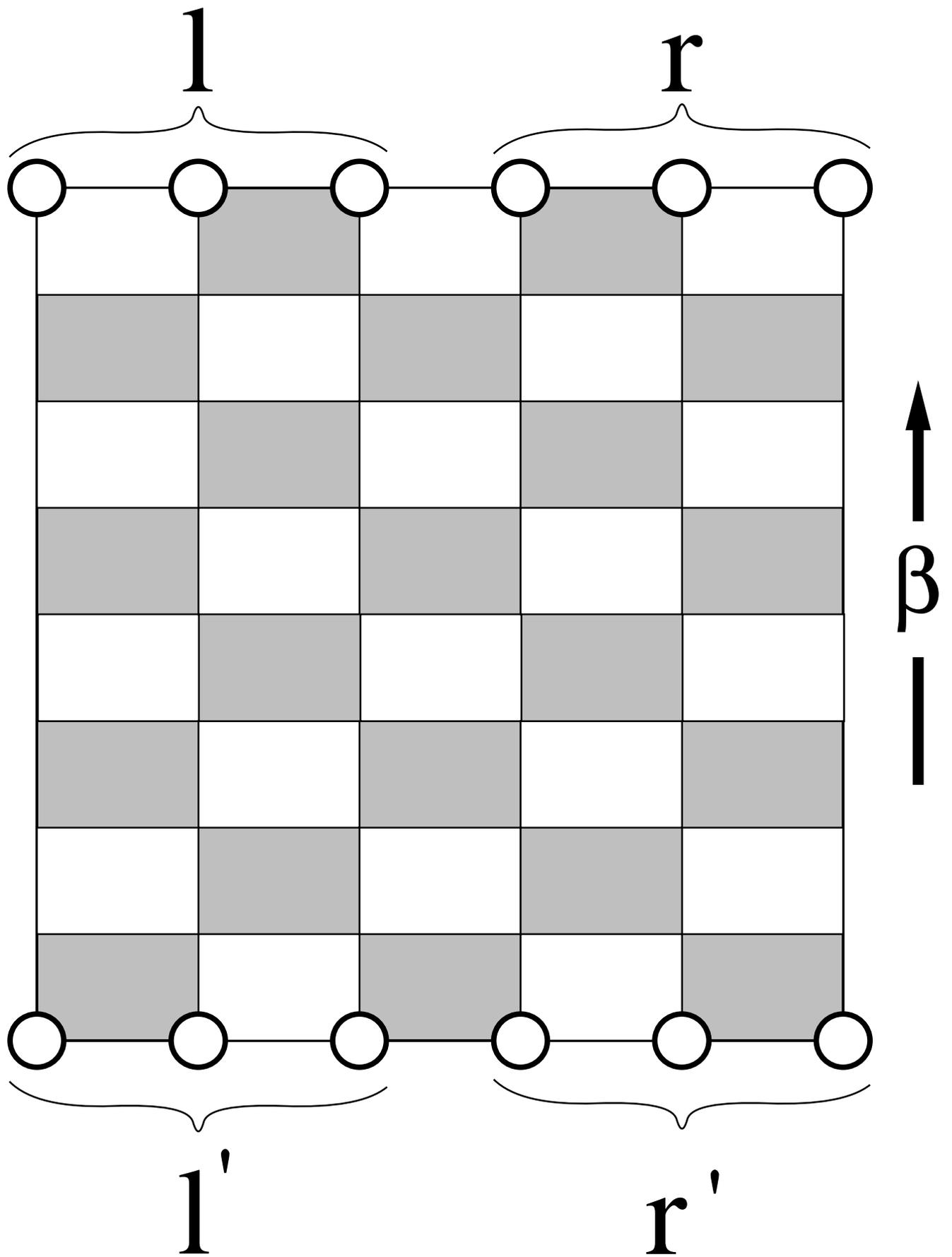
\item[\bf Fig.~3.] Trotter-Suzuki decomposition of the density matrix in Eq.(8).
The label $l$, $r$, $l'$ and $r'$ denote the boundary spin configurations.

\item[\bf Fig.~4.] The density matrix in the Trotter-Suzuki formula Eq.(8): (a)
partition function ${\rm Tr} \, \rho$. (b) density sub matrix $\rho^L_{l l'}$.

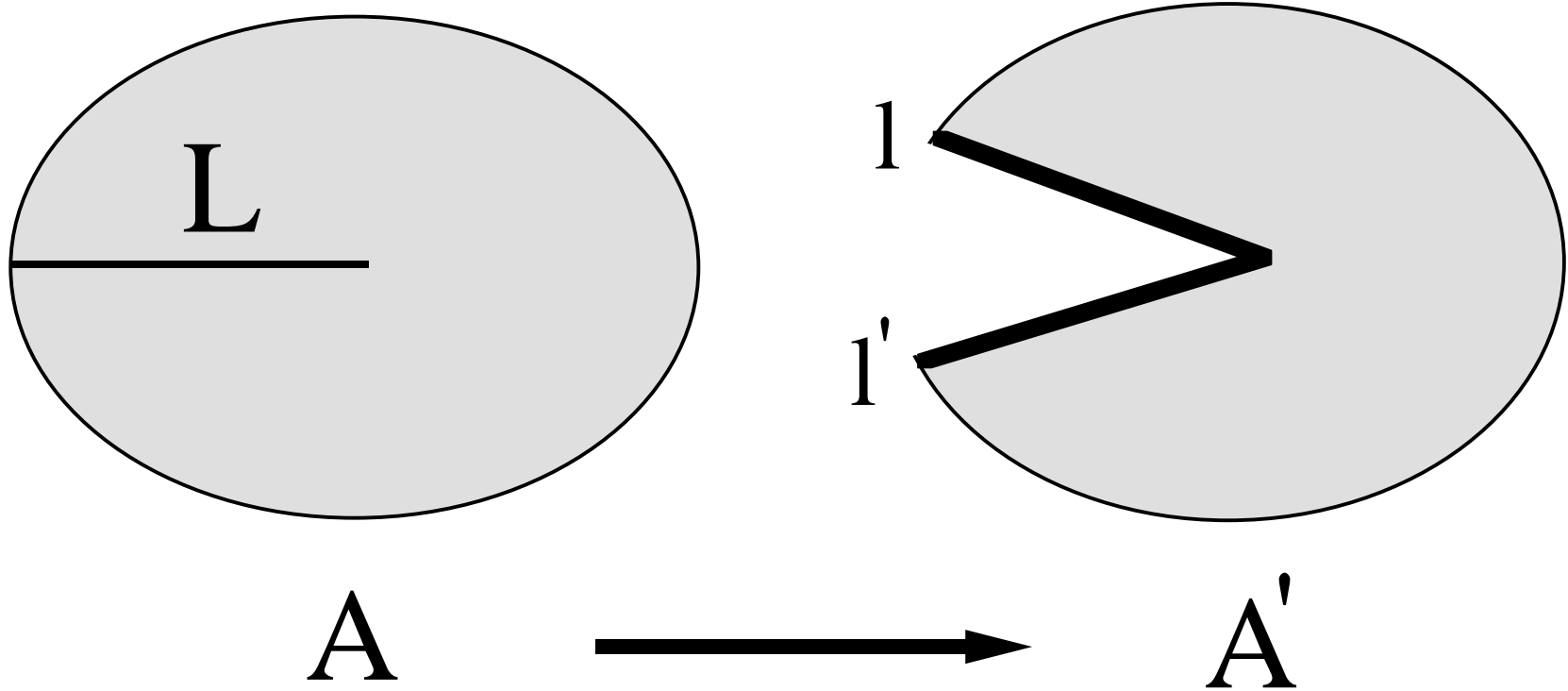
\item[\bf Fig.~5.] We derive the DSM of the finite size system
[$A$] by cutting it along the curve $L$. Compared to [$A$], the system [$A'$]
has additional boundaries $l$ and $l'$.

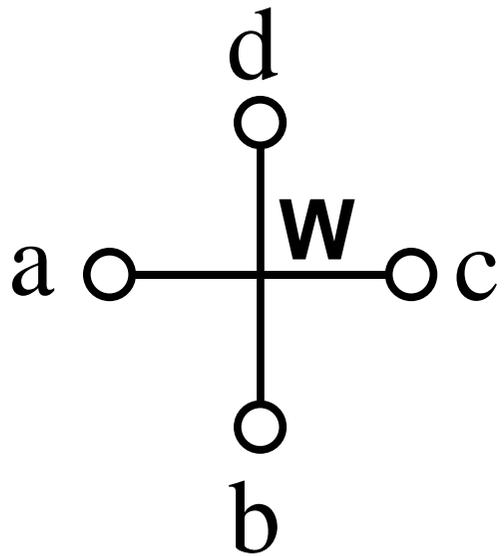
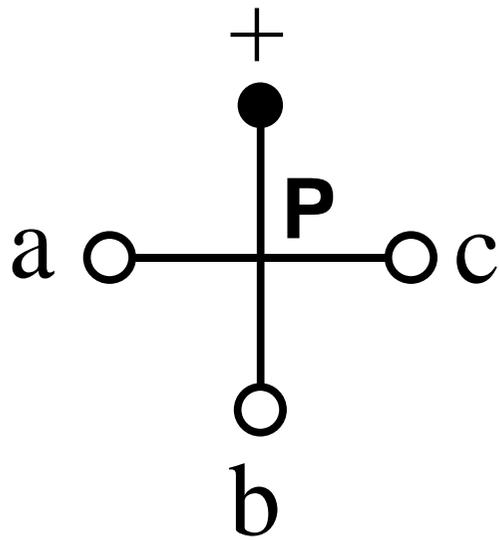
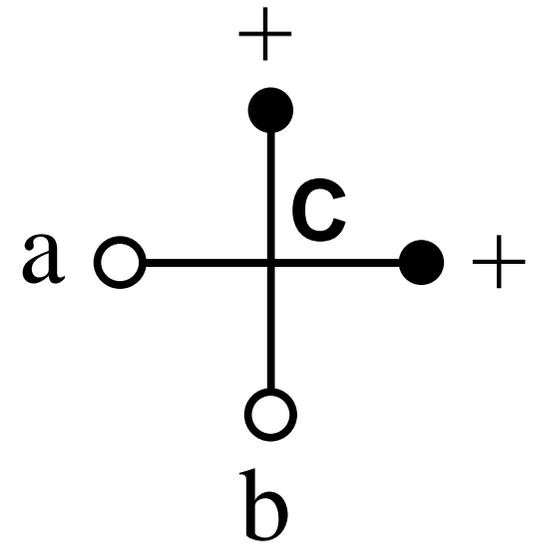
\item[\bf Fig.~6.] Boltzmann weights of the symmetric 16-vertex model; $P$
and $C$ are the boundary weights in Eq.(10) and Eq.(11), respectively.

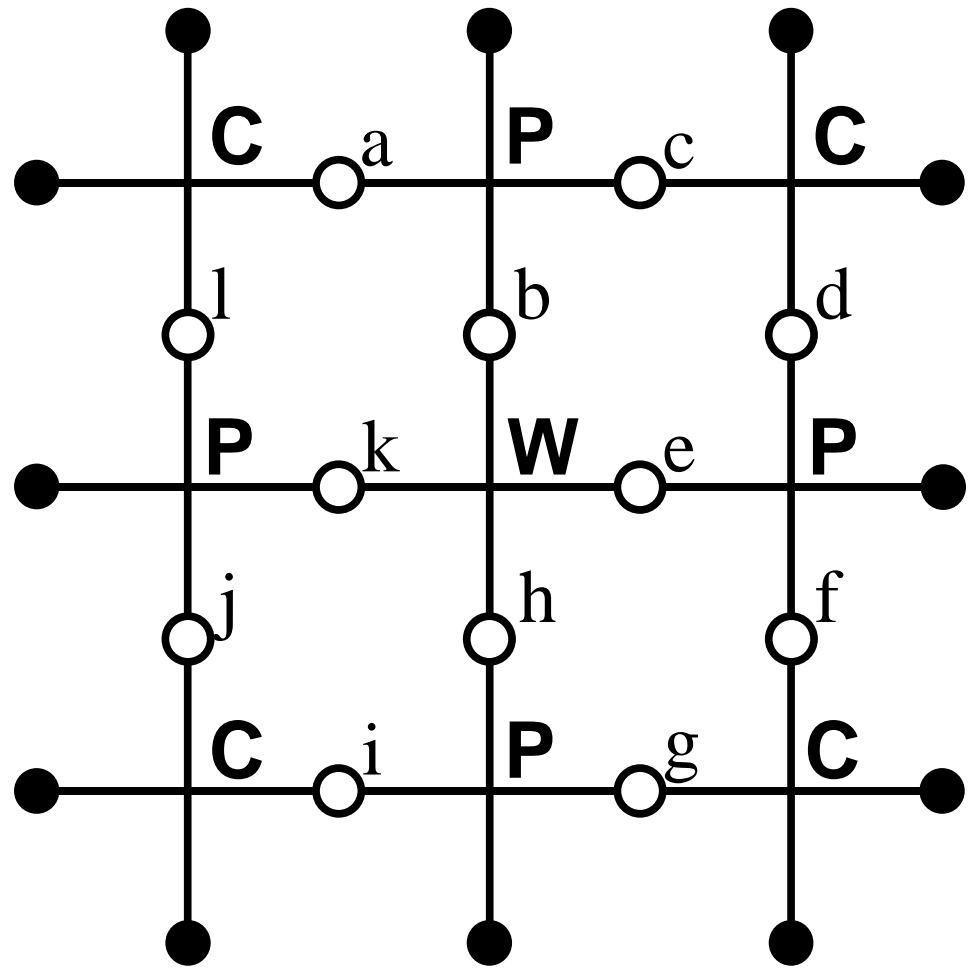
\item[\bf Fig.~7.] A square system of the linear dimension $2N+1 = 3$, whose
partition function is given by Eq.(12).

\item[\bf fig.~8.] Recursive definition of (a) $P^N$ in Eq.(13) and (b) $C^N$ in
Eq.(15). The shown examples are $P^3$ and $C^2$.

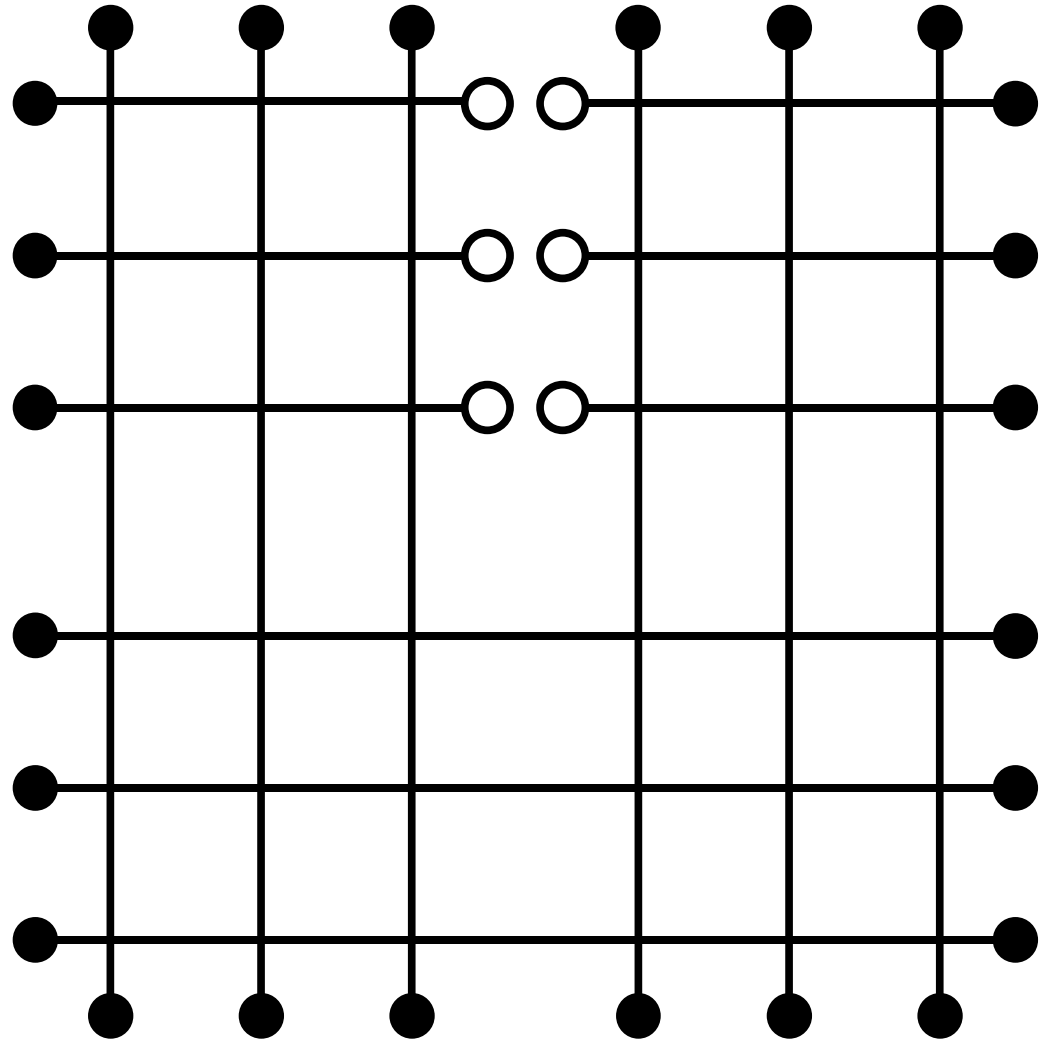
\item[\bf fig.~9.] The density submatrix $\rho^L$ of the square system [$A$] is
expressed as a product of four CTMs. (Eq.(16).)

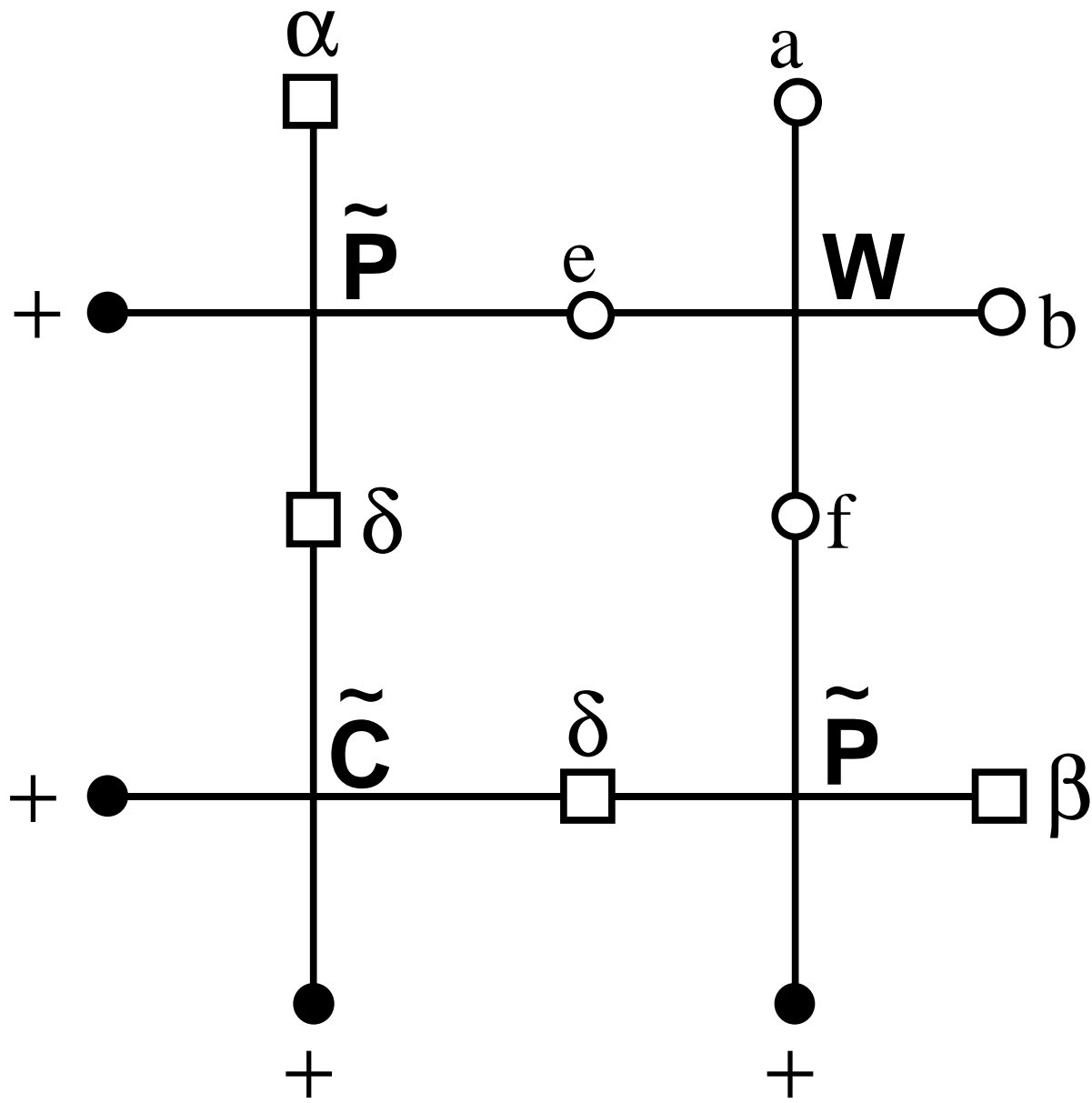
\item[\bf Fig.~10] Area extension of the renormalized CTM in Eq.(20).

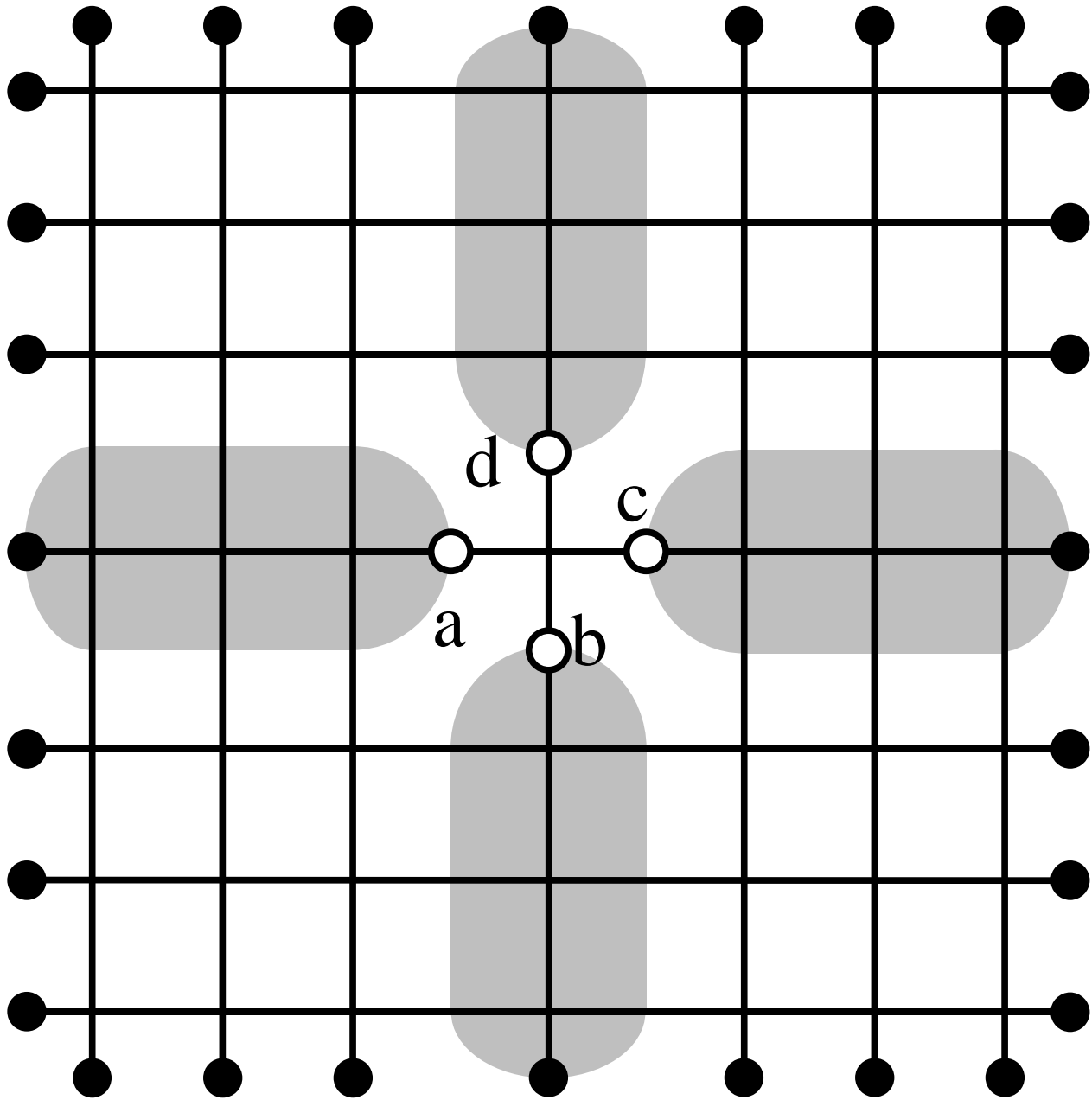
\item[\bf Fig.~11.] A square system of size $2N+1$ is expressed as a product
among ${\tilde C}^N$, ${\tilde P}^N_{a \, (bcd)}$, and $W$.

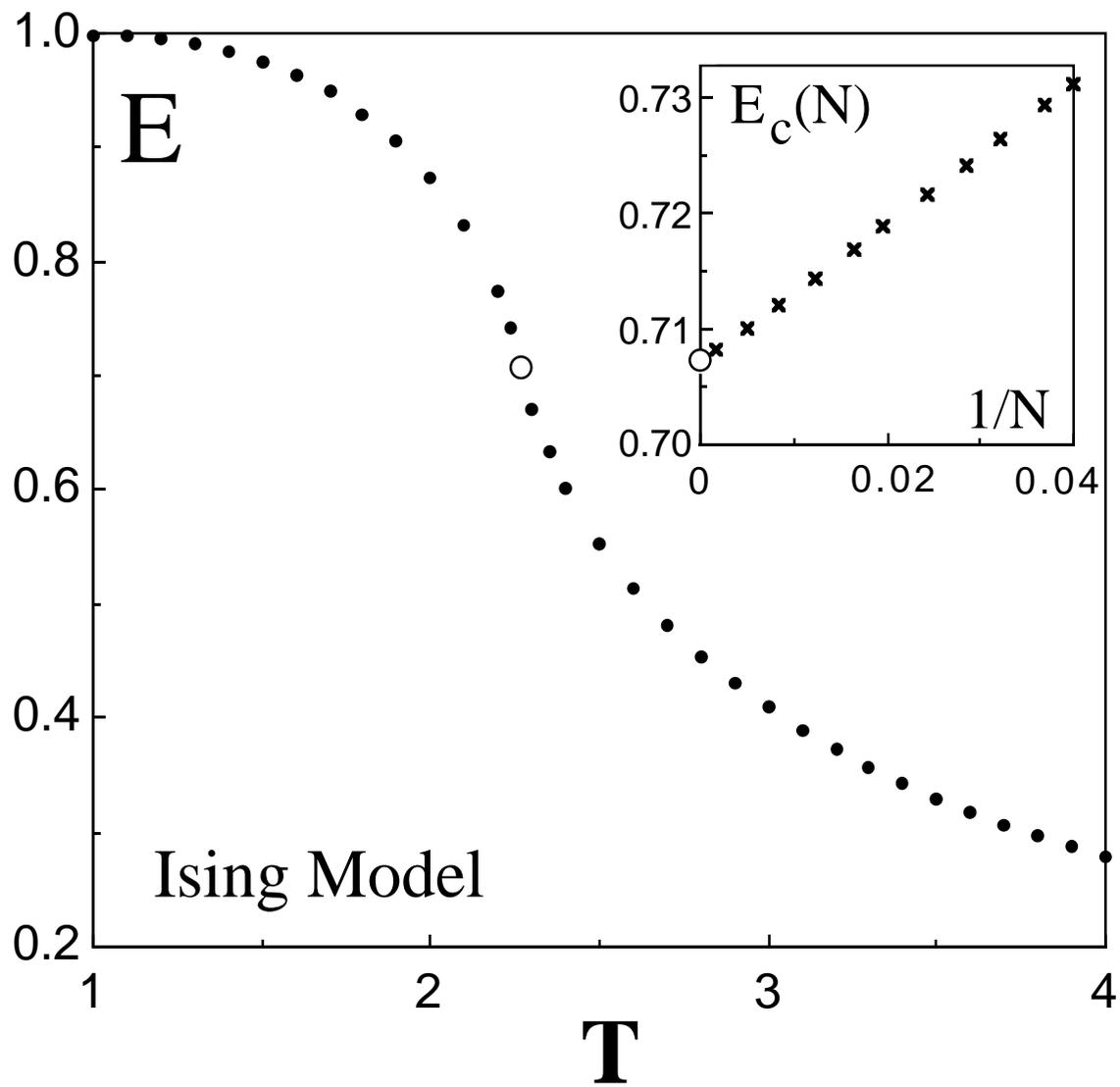
\item[\bf Fig.~12.] Local Energy $E(\infty)$ of the Ising Model at
the center of the square system. The inset shows the size dependence of $E(N)$
at the critical temperature $T_c$.

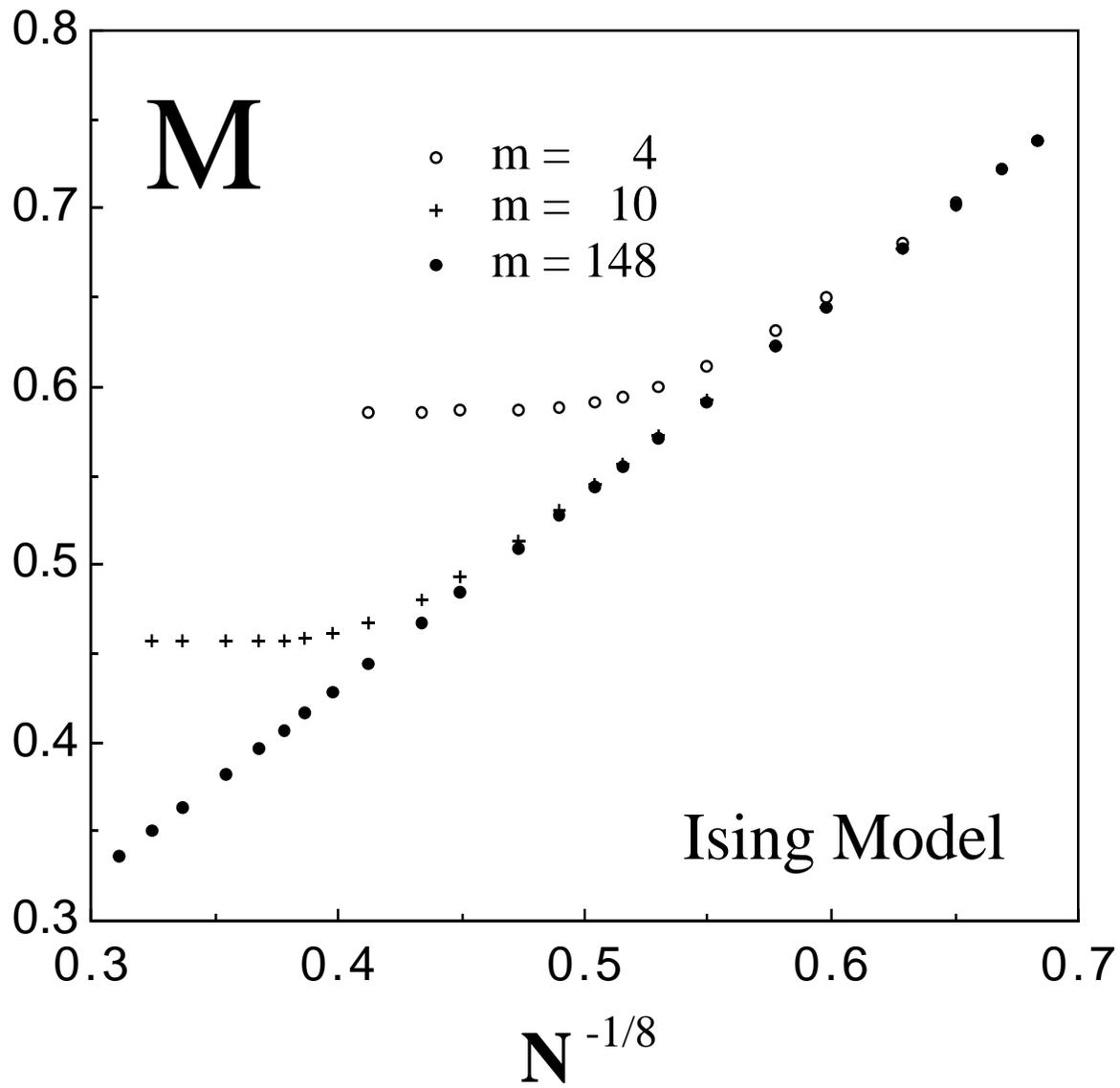
\item[\bf Fig.~13] Local Magnetization of the Ising Model at $T_c$.

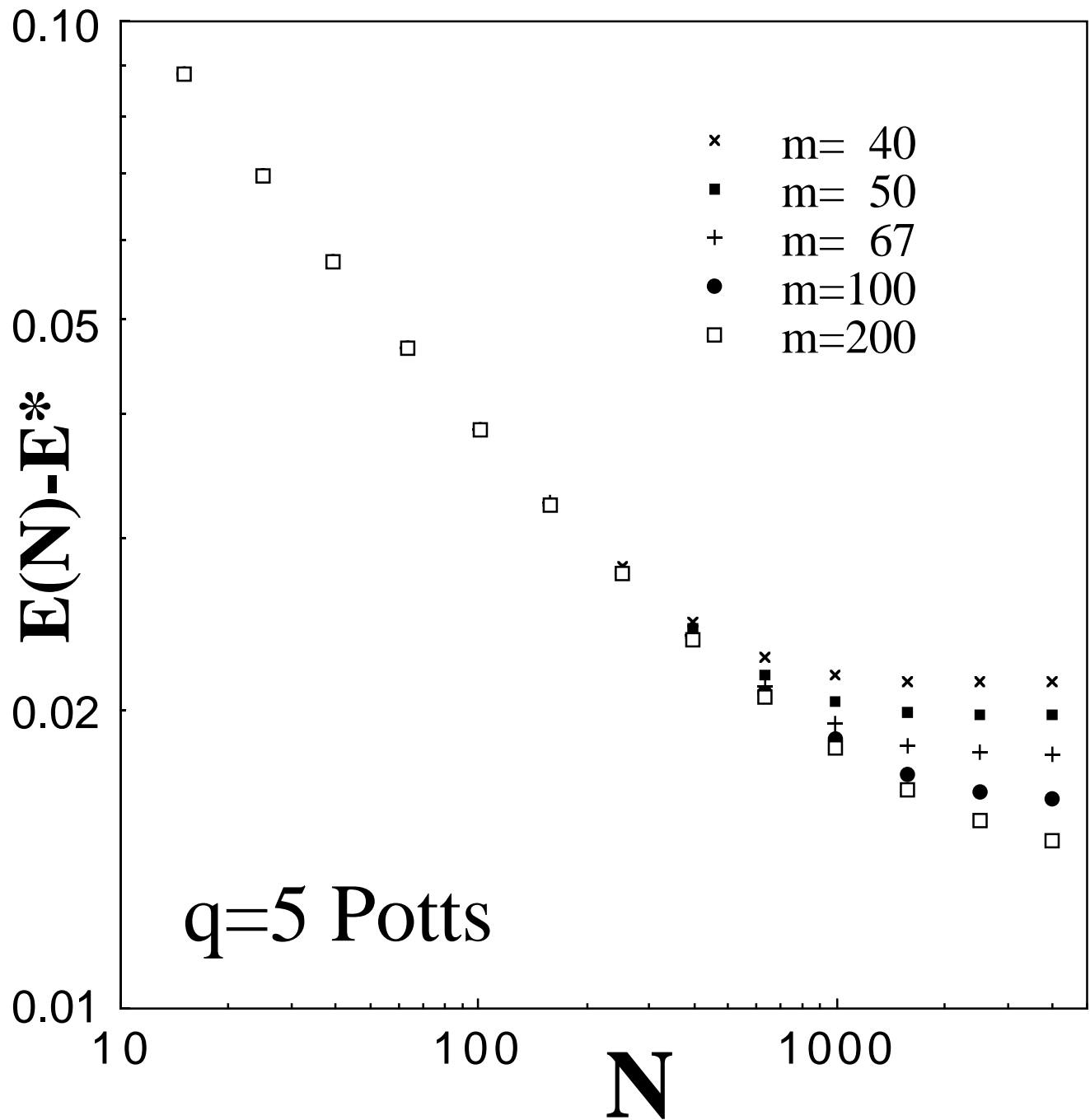
\item[\bf Fig.~14] Local Energy of the $q$=$5$ Potts model.

\end{itemize}
\end{document}